\documentclass[prl,aps,twocolumn,showpacs,floatfix,superscriptaddress]{revtex4}
\usepackage{graphicx}
\usepackage{amsmath}
\usepackage{amssymb}

\begin{document}

\title{Equilibration of Luttinger liquid and conductance of quantum wires}

\author{K. A. Matveev} 

\affiliation{Materials Science Division, Argonne National Laboratory,
  Argonne, IL 60439, USA}

\author{A. V. Andreev} 

\affiliation{Department of Physics, University of Washington, Seattle, WA
  98195, USA}

\date{December 3, 2010}

\begin{abstract}

  Luttinger liquid theory describes one-dimensional electron systems in
  terms of non-interacting bosonic excitations.  In this approximation
  thermal excitations are decoupled from the current flowing through a
  quantum wire, and the conductance is quantized.  We show that relaxation
  processes not captured by the Luttinger liquid theory lead to
  equilibration of the excitations with the current and give rise to a
  temperature-dependent correction to the conductance.  In long wires, the
  magnitude of the correction is expressed in terms of the velocities of
  bosonic excitations.  In shorter wires it is controlled by the
  relaxation rate.

\end{abstract}


\maketitle

The conductance of quantum wires measured at low temperatures is quantized
in units of $2e^2/h$ \cite{old}.  This universality is commonly
interpreted in the framework of the Luttinger liquid theory \cite{haldane,
  giamarchi}, according to which the conductance is not affected by either
the electron-electron interactions in the wire or the temperature
\cite{maslov}.  On the other hand, recent experiments show
temperature-dependent corrections to the conductance \cite{thomas,
  kristensen, reilly, cronenwett, crook}, in apparent contradiction with
the theory \cite{maslov}.  While the origin of the corrections remains the
subject of debate \cite{7theory}, much of the recent theoretical interest
focused on the physics of one-dimensional fermions not captured by the
Luttinger liquid theory \cite{pustilnik, meyer, lunde, rech, micklitz}.

The concept of the Luttinger liquid applies to generic low-energy
properties of systems of interacting one-dimensional fermions.  The key
feature of this theory is that the elementary excitations can be viewed as
non-interacting bosons with an acoustic spectrum:
\begin{equation}
  \label{eq:Luttinger_Hamiltonian}
  H=\sum_q \hbar v|q|b_q^\dagger b_q^{} + 
           \frac{\pi\hbar}{2L}[v_N (N-N_0)^2+v_J J^2],
\end{equation}
Ref. \cite{haldane}.  Here $b_q$ is the operator destroying a boson with
momentum $\hbar q$, the velocities $v$, $v_N$, and $v_J$ may depend on the
electron-electron interactions, $L$ is the system size (periodic boundary
conditions are assumed).  In addition to the boson occupation numbers
$N_q$, the state of the system is described by two integers, $N$ and $J$.
The latter can be related to the total numbers of the right- and
left-moving electrons in the system, $J=N^R-N^L$, and $N=N^R+N^L$; $N_0$
is the total number of electrons in some reference state.

It is important to note that within the Luttinger liquid approximation the
numbers $N^R$ and $N^L$ of the right- and left-movers are conserved, i.e.,
electron-electron interactions do not cause backscattering of electrons.
As a result, if the interacting quantum wire coupled to non-interacting
leads is described by the Luttinger liquid model \cite{maslov}, the
electrons entering the wire from the leads always pass through it, and the
interactions do not affect the conductance.  On the other hand, if the
interactions are treated as a small perturbation to the non-interacting
fermion picture, they do lead to backscattering of electrons and reduce
the conductance \cite{lunde,rech,micklitz}.  The backscattering occurs via
three-electron scattering processes, which are not included in the
Luttinger liquid theory.

The corrections to the quantized conductance \cite{thomas, kristensen,
  reilly, cronenwett, crook} are observed in the regime of low electron
density, where the electron-electron interactions are strong.  Thus the
theory \cite{rech,micklitz} based on the picture of weakly interacting
electrons cannot be applied to the experiments directly, and the
generalization to the case of arbitrary interaction strength is needed.
In this paper we develop such a theory using the Luttinger liquid picture
as a starting point and accounting for the electron backscattering
processes.

We start by noting that the Luttinger liquid Hamiltonian
(\ref{eq:Luttinger_Hamiltonian}) describes the stable low-energy fixed
point of a one-dimensional system of interacting fermions.  The full
low-energy Hamiltonian consists of (\ref{eq:Luttinger_Hamiltonian}) and a
number of perturbations describing interactions between the excitations.
Although formally irrelevant, these terms are responsible for establishing
the thermal equilibrium in the system.  The equilibrium distribution of
the bosonic excitations is dictated by conservation laws and independent
of the detailed form of the irrelevant perturbations,
\begin{equation}
  \label{eq:Boson_distribution}
  N_q=\frac{1}{e^{\hbar(v|q|-uq)/T}-1}.
\end{equation}
Due to the conservation of energy and momentum, the distribution function
is characterized by two parameters, temperature $T$ and velocity $u$.
Relaxation towards the equilibrium distribution
(\ref{eq:Boson_distribution}) is due to the irrelevant perturbations which
have the form of higher-order terms in bosonic variables, and the
respective relaxation time is expected to scale as a power of temperature,
$\tau_0\propto T^{-\alpha}$.

The boson distribution (\ref{eq:Boson_distribution}) can be derived from
the general principle that for systems where energy and momentum are
conserved the probability of realization of a given microscopic state $i$
depends on its energy $E_i$ and momentum $P_i$ as $e^{-(E_i-uP_i)/T}$.
The momentum of a Luttinger liquid is given by \cite{haldane}
\begin{equation}
  \label{eq:Luttinger_Momentum}
  P=p_F J +\sum_q \hbar q\, b_q^\dagger b_q^{},
\end{equation}
where $p_F=\pi\hbar N/L$ and the first term represents the momentum of a
filled Fermi surface with $J=N^R-N^L$ extra electrons at the right Fermi
point.  The presence of $J$ in the momentum (\ref{eq:Luttinger_Momentum})
means that its distribution $e^{-(\Delta J^2-up_FJ)/T}$ and the boson
occupation numbers (\ref{eq:Boson_distribution}) are controlled by the
same parameter $u$. (Here $\Delta=\pi\hbar v_J/2L$.)  In particular, the
thermodynamic average of $J$ is given by $J=Nu/v_J$.  Taking into account
the relation \cite{haldane} between $J$ and the electric current,
$I=ev_J(J/L)$, we conclude that at full equilibrium the velocity $u$ of
the boson gas coincides with the drift velocity
\begin{equation}
  \label{eq:v_d}
  v_d=\frac{I}{e}\,\frac{L}{N}.
\end{equation}
The conclusion $u=v_d$ is obvious in the presence of Galilean invariance,
where the boson gas should be at rest in a frame moving with the drift
velocity.

Establishment of full thermodynamic equilibrium requires the presence of
backscattering processes which change the numbers of the right- and
left-moving electrons $N^R$ and $N^L$ and transfer the corresponding
momentum to the boson excitations.  Such processes are not accounted for
in the Luttinger liquid theory and, as we discuss below, occur at a time
scale $\tau$ much longer than the time $\tau_0$ required for the formation
of the boson distribution (\ref{eq:Boson_distribution}).  Therefore, at
time scales $t\gg \tau_0$ equilibration processes may be described by
relaxation of the velocity $u$ of the boson gas towards $v_d$,
\begin{equation}
  \label{eq:relaxation}
  \dot u = -\frac{u-v_d}{\tau}.
\end{equation}
Due to the time dependence of $u$, the total momentum of the bosons $P_b$
changes as well.  The rate of this change is easily obtained from
Eqs.~(\ref{eq:Boson_distribution}) and (\ref{eq:Luttinger_Momentum}),
$\dot P_b=(\pi LT^2/3\hbar v^3)\dot u$.  Conservation of the total
momentum (\ref{eq:Luttinger_Momentum}) of the Luttinger liquid implies
that the number of right-moving electrons changes at the rate $\dot
N^R=\dot J/2=-\dot P_b/2p_F$, resulting in
\begin{equation}
  \label{eq:NRdot}
  \frac{\dot N^R}{L} = \frac{\pi}{6\hbar}\,
                       \frac{T^2}{v^3p_F}\,
                       \frac{u-v_d}{\tau}.
\end{equation}
Equations (\ref{eq:relaxation}) and (\ref{eq:NRdot}) describe
equilibration of a system of interacting electrons at times $t\gg \tau_0$
in terms of the phenomenological parameter $\tau$.

In the case of weak interactions the backscattering processes involve a
hole diffusing through the bottom of the band due the three-particle
collisions \cite{micklitz}.  Because of small occupation probabilities of
the hole states, the backscattering rate follows the Arrhenius law,
$\tau^{-1}\propto e^{-D/T}$, with the activation energy $D$ given by the
Fermi energy.  The momentum of the backscattered particle is redistributed
among the particle-hole pairs near the Fermi points, which correspond to
the bosons in the Luttinger liquid theory.  In the case of strong
interactions, the one-dimensional electrons form a Wigner crystal.  The
excitations of the crystal are phonons.  At low energies they have 
linear dispersion and correspond to the bosons in the Luttinger liquid.
The total (quasi-) momentum of the phonons can change via umklapp
processes, whose rate also follows the Arrhenius law, with the activation
energy $D$ determined by the Debye frequency \cite{equilibrationWigner}.

The above examples demonstrate that the backscattering processes involve
states at energies of the order of the bandwidth.  Such processes are
neglected in the renormalization group schemes describing the low-energy
properties of Luttinger liquids.  Although at arbitrary interaction
strengths these processes have not been studied microscopically, we expect
the scattering rate to follow the Arrhenius law with the activation energy
of the order of the bandwidth, $D\sim vp_F$.

In the following we consider the conductance of a long uniform quantum
wire adiabatically connected to ideal metal leads, starting with the case
of spinless electrons.  The interacting electrons in the wire form a
Luttinger liquid, whereas the leads are assumed to be non-interacting.
The numbers of the right- and left-moving electrons in the wire are
controlled by the chemical potentials of the leads.  In the case of a
relatively short wire, $L\ll v\tau$, the time of flight of an electron
across the wire is shorter than $\tau$, and backscattering is negligible.
In this case the conductance is $G_0=e^2/h$.  In longer wires, $L\gtrsim
v\tau$, a significant fraction of right-movers are backscattered, and the
electric current $I$ through the wire is reduced,
\begin{equation}
  \label{eq:current}
  I=G_0V + e\dot N^R.
\end{equation}
Here $V$ is the bias voltage.  Equation (\ref{eq:current}) expresses the
particle number conservation in the system.  In the case of weakly
interacting electrons it was derived in Ref.~\cite{rech}.  Since $N^R$ and
$N^L$ retain their meanings as the numbers of right- and left-moving
electrons in the Luttinger liquid, the derivation of Ref.~\cite{rech}
applies at any interaction strength.

We now use conservation of momentum and energy to find $\dot N^R$ in
Eq.~(\ref{eq:current}) in the linear response regime $I\to0$.
Conservation of the momentum (\ref{eq:Luttinger_Momentum}) implies that
backscattering of a single right-mover, $\Delta N^R=-1$, is accompanied by
transfer of momentum $2p_F$ to the bosons.  On the other hand, according
to Eq.~(\ref{eq:Luttinger_Hamiltonian}) at $J\to0$ the total energy $E$ of
the bosons should remain unchanged.  Thus, the momenta transferred to the
right- and left-moving bosons must be equal, i.e., $\Delta P^{R,L}=p_F$
and $\Delta E^{R,L}=\pm vp_F$.  We therefore conclude that backscattering
of right-moving electrons changes the energy of the right-moving bosons at
a rate
\begin{equation}
  \label{eq:ERdot}
  \dot E^R=-vp_F \dot N^R.
\end{equation}
In the absence of backscattering all the bosons in the system are in
thermal equilibrium with the leads, whose temperatures are assumed to be
equal.  Thus the energy current in the wire vanishes.  Backscattering
transfers energy from the left- to the right-moving bosons at the rate
(\ref{eq:ERdot}), resulting in the net energy current
\begin{equation}
  \label{eq:jE}
  j_E=\dot E^R.
\end{equation}
Equations (\ref{eq:current})--(\ref{eq:jE}) express conservation of
particle number, momentum, and energy in the system.  To make further
progress we need to discuss the boson distribution function in a wire
connected to leads.

As argued above, an appreciable correction to the conductance arises when
the wire is rather long, $L\gtrsim v\tau$.  The exponentially long
relaxation time $\tau$ describes establishment of the full equilibrium in
the system.  On the other hand, equilibration of the bosons to the
distribution (\ref{eq:Boson_distribution}) occurs at the much faster time
scale $\tau_0$.  Therefore at $L\gtrsim v\tau\gg v\tau_0$ the boson
distribution inside the wire takes the form (\ref{eq:Boson_distribution})
with velocity $u$ that may or may not satisfy the full equilibrium
condition $u=v_d$.  In the linear response regime $J\to0$ only the bosons
contribute to the energy current in the Luttinger liquid.  Its evaluation
for the distribution (\ref{eq:Boson_distribution}) is straightforward and
gives
\begin{equation}
  \label{eq:jEu}
  j_E=\frac\pi3\, \frac{T^2}{\hbar v}\,u.
\end{equation}
The parameters $u$ and $T$ of the boson distribution
(\ref{eq:Boson_distribution}) may depend on position $x$ along the wire.
In the linear regime, $I\to0$, one expects $u\propto I$ and $dT/dx\propto
I$.  Then from conservation of energy, $dj_E/dx=0$, and Eq.~(\ref{eq:jEu})
we conclude that $u$ is position independent.

Combining Eqs.~(\ref{eq:ERdot})-(\ref{eq:jEu}) we find the following
relation between the backscattering rate $\dot N^R$ and velocity $u$,
\begin{equation}
  \label{eq:NRdot-u}
  \dot N^R = - \frac{\pi}{3}\, \frac{T^2}{\hbar v^2 p_F}\, u.
\end{equation}
This equation expresses the energy and momentum conservation laws in a
wire connected to ideal leads and must be satisfied regardless of the
equilibration rate $\tau$.  On the other hand, if the equilibration rate
is known, the backscattering rate $\dot N^R$ can also be related to the
boson gas velocity $u$ via Eq.~(\ref{eq:NRdot}).  Comparing
(\ref{eq:NRdot}) and (\ref{eq:NRdot-u}) we conclude that in a wire of
length $L$ the boson gas velocity $u$ is related to the drift velocity
(\ref{eq:v_d}) as
\begin{equation}
  \label{eq:u-v_d}
  u=\frac{L}{l_{\rm eq}+L}\,v_d,
\quad
  l_{\rm eq}=2v\tau.
\end{equation}

Substituting Eqs.~(\ref{eq:u-v_d}) and (\ref{eq:NRdot-u}) into
(\ref{eq:current}), we find the conductance of the wire
\begin{equation}
  \label{eq:conductance}
  G=G_0\left(1-\frac{\pi^2}{3}\,
     \frac{T^2}{v^2p_F^2}\frac{L}{l_{\rm eq}+L}\right).
\end{equation}
The second term in Eq.~(\ref{eq:conductance}) represents a correction to
the Luttinger liquid result \cite{maslov} originating from equilibration
of bosonic excitations.  The result (\ref{eq:conductance}) is valid for
any interaction strength.  At weak interactions $v$ coincides with the
Fermi velocity $v_F=p_F/m$, and Eq.~(\ref{eq:conductance}) agrees with the
results of Refs.~\cite{rech,micklitz}.  In short wires, $L\ll l_{\rm eq}$,
the correction to the quantized conductance is proportional to the
equilibration rate $\tau^{-1}\propto e^{-D/T}$ and the length of the wire,
$\delta G\propto L e^{-D/T}$.  As the length of the wire increases, $L\gg
l_{\rm eq}$, the correction saturates at $\delta G\sim (e^2/h)(T/vp_F)^2$,
but remains small within the range of applicability of our theory, $T\ll
vp_F$.  Interestingly, the magnitude of the correction at $L\to \infty$ is
fully characterized by the parameters of the Luttinger liquid Hamiltonian
(\ref{eq:Luttinger_Hamiltonian}), with interaction strength entering via
the speed of the bosonic excitations $v$.

Our result (\ref{eq:conductance}) applies to systems where electron spins
are polarized by an external magnetic field.  On the other hand, the
temperature-dependent corrections to conductance \cite{thomas, kristensen,
  reilly, cronenwett, crook} are observed in unpolarized wires, and spins
are believed to play a significant role in this phenomenon.  We now
generalize our discussion to the case of electrons with spin.  Compared to
Eq.~(\ref{eq:Luttinger_Hamiltonian}), the Hamiltonian of the Luttinger
liquid with spins \cite{giamarchi} has twice the number of modes.  Most
importantly, there are two branches of bosonic excitations, corresponding
to waves of charge and spin densities, and propagating at different
velocities, $v_\rho$ and $v_\sigma$, respectively.  In addition, instead
of the numbers of right- and left-moving electrons $N^{R,L}=(N\pm J)/2$,
the state of each system is described by the numbers of such electrons for
each spin direction, $N^{R,L}_{\uparrow}$ and $N^{R,L}_{\downarrow}$.

The discussion leading to Eq.~(\ref{eq:conductance}) can then be readily
modified as follows.  The conductance $G_0$ of non-interacting wire in
Eq.~(\ref{eq:current}) doubles in the presence of spins.  The number of
right-movers $N^R$ in Eqs.~(\ref{eq:NRdot}), (\ref{eq:current}),
(\ref{eq:ERdot}), and (\ref{eq:NRdot-u}) should be understood as
$N^R_\uparrow+N^R_\downarrow$, whereas $E^R$ in Eqs.~(\ref{eq:ERdot}) and
(\ref{eq:jE}) becomes the total energy of right moving bosons, in both
charge and spin channels.  In the right-hand sides of Eq.~(\ref{eq:NRdot})
and (\ref{eq:jEu}) one has to account for two types of boson excitations
by replacing $v^{-n}\to v^{-n}_\rho + v^{-n}_\sigma$.  The most
significant revision appears in Eq.~(\ref{eq:ERdot}).  When an electron is
backscattered and momentum $p_F$ is transferred to right-moving bosons, it
is distributed unevenly between the charge and spin branches.  Due to the
relatively fast equilibration of bosons their distribution functions have
the form (\ref{eq:Boson_distribution}) with different velocities $v_\rho$
and $v_\sigma$ in the two channels but the same $u$.  Momentum change is
accommodated by a small change of $u$.  As a result, the fraction of the
momentum $p_F$ transferred into charge or spin branch is proportional to
$v_{\rho,\sigma}^{-3}$, whereas the corresponding energies scale as
$v_{\rho,\sigma}^{-2}$.  We thus generalize Eq.~(\ref{eq:ERdot}) as
\footnote{We note that in the presence of spins $p_F=\pi\hbar N/2L$.}
\begin{equation}
  \label{eq:ERdot-spins}
  \dot E^R =
  -\frac{v_\rho^{-2}+v_\sigma^{-2}}{v_\rho^{-3}+v_\sigma^{-3}}\,
  p_F \dot N^R.
\end{equation}
As a result the equilibration length $l_{\rm eq}$ in Eq.~(\ref{eq:u-v_d})
becomes
 \begin{equation}
   \label{eq:l_eq_spins}
   l_{\rm eq}=2\tau\frac{v_\rho^{-1}+v_\sigma^{-1}}
                  {v_\rho^{-2}+v_\sigma^{-2}},
 \end{equation}
 and we recover the general form (\ref{eq:conductance}) of the dependence
 of conductance on the length of the wire with the following
 modifications:
\begin{equation}
  \label{eq:spin-modifications}
  G_0\to\frac{2e^2}{h},
\quad
  v^2\to\frac{2(v_\rho^{-2}+v_\sigma^{-2})}
             {(v_\rho^{-1}+v_\sigma^{-1})(v_\rho^{-3}+v_\sigma^{-3})}.
\end{equation}
The presence of both $v_\rho$ and $v_\sigma$ in the expression for the
correction to the quantized conductance $2e^2/h$ reflects the fact that
the momentum of backscattered electrons is distributed among the bosonic
excitations in both the charge and spin channels.

The two velocities $v_\rho$ and $v_\sigma$ coincide with the Fermi
velocity $v_F$ in the limit of weak interactions.  For repulsive
interactions, $v_\rho>v_F$ and $v_\sigma<v_F$.
The experimentally observed temperature-dependent corrections to the
conductance of quantum wires \cite{thomas, kristensen, reilly, cronenwett,
  crook} are most prominent at low electron density.  In this regime the
Coulomb interactions between electrons are effectively very strong,
$e^2/\hbar v_F\gg1$, and $v_\rho\gg v_\sigma$.  The expression for the
conductance of a quantum wire given by Eqs.~(\ref{eq:conductance}) and
(\ref{eq:spin-modifications}) then simplifies,
\begin{equation}
  \label{eq:conductance-strong-interactions}
  G=\frac{2e^2}{h}\left[1-\frac{\pi^2}{6}\,
     \left(\frac{T}{v_\sigma p_F}\right)^2\frac{L}{l_{\rm eq}+L}\right],
\end{equation}
with $l_{\rm eq}=2v_\sigma \tau$, see Eq.~(\ref{eq:l_eq_spins}).
Importantly, the magnitude of the correction to the quantized conductance
is controlled by the spin velocity.  The reason is that at $v_\sigma\ll
v_\rho$ both the energy and momentum of the bosonic excitations are
dominated by the softer spin branch.

Similarly to the case of spinless electrons, for long wires, $L\gg l_{\rm
  eq}$, the correction shows quadratic temperature dependence, $\delta G
\sim (e^2/h) (T/v_\sigma p_F)^2$.  Within the range of applicability of
our theory, $T\ll v_\sigma p_F$, the correction remains small.  In short
wires, $L\ll l_{\rm eq}$, the temperature dependence of $\delta G$ is
controlled by that of the relaxation time, $\delta G\propto \tau^{-1}$.
We are not aware of microscopic calculations of the relaxation time $\tau$
for strongly interacting electrons with spin.  However, based on the
analogy with the case of spinless electrons \cite{equilibrationWigner}, we
expect activated temperature dependence $\tau^{-1}\propto
e^{-D_\sigma/T}$, with the activation energy of the order of the bandwidth
of the spin excitations, $D_\sigma\sim v_\sigma p_F\ll v_\rho p_F$.

In a wire of fixed length the crossover between the regimes of long and
short wires can be explored by changing the temperature $T$.  Because of
the exponential temperature dependence of $\tau$, the condition $L\sim
2v_\sigma\tau$ is satisfied at $T\sim D_\sigma/\ln N$, where $N$ is the
number of electrons in the one-dimensional part of the device.  At $T\ll
D_\sigma/\ln N$ the activated behavior $\delta G\propto e^{-D_\sigma/T}$
should be observed, in agreement with the experiment \cite{kristensen}.
The quadratic temperature dependence is expected in the relatively narrow
range $D_\sigma/\ln N\ll T\ll D_\sigma$, and its experimental observation
requires working with rather long wires.  At the upper limit of this range
$\delta G\sim e^2/h$.  At this point the system crosses over into the
so-called spin-incoherent regime.  Study of this crossover would be very
interesting from both the theoretical point of view and because of its
possible relevance for the experiments \cite{thomas, kristensen, reilly,
  cronenwett, crook} showing a shoulder-like feature in the conductance at
$0.7\times 2e^2/h$.

To summarize, we developed a theory of conductance of quantum wires with
strongly interacting electrons.  Our theory extends the Luttinger
liquid picture to account for the equilibration processes.  Such processes
involve backscattering of electrons and result in a temperature-dependent
correction to quantized conductance of the wire.

The authors are grateful to T. Micklitz, M. Pustilnik, and J. Rech, for
discussions.  This work was supported by the U.S. Department of Energy
under Contract Nos. DE-AC02-06CH11357 and DE-FG02-07ER46452.

\end{document}